\documentclass{emulateapj}\usepackage{apjfonts}
\pdfoutput=1
\usepackage{natbib}
\newcommand{\target}{HD 131835}
\newcommand{\um}{$\mu$m}

\begin{document}
% \journalinfo{}

\submitted{Accepted for publication in ApJL}

% title and author info
\title{FIRST SCATTERED-LIGHT IMAGE OF THE DEBRIS DISK AROUND HD 131835 WITH THE GEMINI PLANET IMAGER}
% \shorttitle{}
\author{
Li-Wei Hung\altaffilmark{1},
Gaspard Duchene\altaffilmark{2,3,4},
Pauline Arriaga\altaffilmark{1},
Michael P. Fitzgerald\altaffilmark{1},
J\'er\^ome Maire\altaffilmark{5},
Christian Marois\altaffilmark{6,7},
Maxwell A. Millar-Blanchaer\altaffilmark{8},
Sebastian Bruzzone\altaffilmark{9},
Abhijith Rajan\altaffilmark{10},
Laurent Pueyo\altaffilmark{11},
Paul G. Kalas\altaffilmark{2},
Robert J. De Rosa\altaffilmark{2},
James R. Graham\altaffilmark{2}, 
Quinn Konopacky\altaffilmark{12},
Schuyler G. Wolff\altaffilmark{11,13},
S. Mark Ammons\altaffilmark{14},
Christine Chen\altaffilmark{11},
Jeffrey K. Chilcote\altaffilmark{5},
Zachary H. Draper\altaffilmark{7},
Thomas M. Esposito\altaffilmark{1,2},
Benjamin Gerard\altaffilmark{6,7},
Stephen Goodsell\altaffilmark{15,16},
Alexandra Greenbaum\altaffilmark{11,13},
Pascale Hibon\altaffilmark{15},
Sasha Hinkley\altaffilmark{17},
Bruce Macintosh\altaffilmark{14,18}, 
Franck Marchis\altaffilmark{19},
Stanimir Metchev\altaffilmark{9,20},
Eric L. Nielsen\altaffilmark{18, 19},
Rebecca Oppenheimer\altaffilmark{21},
Jenny Patience\altaffilmark{10},
Marshall Perrin\altaffilmark{11},
Fredrik T. Rantakyr\"o\altaffilmark{15},
Anand Sivaramakrishnan\altaffilmark{11},
Jason J. Wang\altaffilmark{2},
Kimberly Ward-Duong\altaffilmark{10},
and Sloane J. Wiktorowicz\altaffilmark{22}
}
\altaffiltext{1}{Dept. of Physics \& Astronomy, UCLA, Los Angeles, CA 90095}
\altaffiltext{2}{Astronomy Dept., Univ. of California, Berkeley CA 94720-3411}
\altaffiltext{3}{Univ. Grenoble Alpes, IPAG, F-38000 Grenoble, France}
\altaffiltext{4}{CNRS, IPAG, F-38000 Grenoble, France}
\altaffiltext{5}{Dunlap Institute for Astronomy \& Astrophysics, University of Toronto, 50 St. George Street, Toronto, Ontario, Canada M5S 3H4}
\altaffiltext{6}{National Research Council of Canada Herzberg, 5071 West Saanich Road, Victoria, BC V9E 2E7, Canada}
\altaffiltext{7}{University of Victoria, 3800 Finnerty Rd, Victoria, BC, V8P 5C2, Canada}
\altaffiltext{8}{Dept. of Astronomy \& Astrophys., Univ. of Toronto, Toronto ON M5S 3H4, Canada}
\altaffiltext{9}{Department of Physics \& Astronomy, Centre for Planetary and Space Exploration, University of Western Ontario, London, ON, N6A 3K7, Canada}
\altaffiltext{10}{School of Earth and Space Exploration, Arizona State Univ., PO Box 871404, Tempe, AZ 85287}
\altaffiltext{11}{Space Telescope Science Institute, 3700 San Martin Drive, Baltimore MD 21218}
\altaffiltext{12}{Center for Astrophysics and Space Sciences, University of California, San Diego, 9500 Gilman Drive, La Jolla, CA 92093, USA}
\altaffiltext{13}{Physics \& Astronomy Dept., Johns Hopkins Univ., Baltimore MD, 21218}
\altaffiltext{14}{Lawrence Livermore National Laboratory, 7000 East Ave., Livermore, CA 94040}
\altaffiltext{15}{Gemini Observatory, Casilla 603, La Serena, Chile}
\altaffiltext{16}{Durham University, Stockton Road, Durham, DH1 3LE, UK}
\altaffiltext{17}{School of Physics, Univ. of Exeter, Stocker Road, Exeter, EX4 4QL, UK}
\altaffiltext{18}{Kavli Institute for Particle Astrophys. \& Cosmology, Stanford Univ., Stanford, CA 94305}
\altaffiltext{19}{SETI Institute, Carl Sagan Center, 189 Bernardo Avenue, Mountain View, CA 94043}
\altaffiltext{20}{Stony Brook University, 100 Nicolls Rd, Stony Brook, NY 11794, USA}
\altaffiltext{21}{American Museum of Natural History, New York, NY 10024}
\altaffiltext{22}{Dept. of Astronomy \& Astrophys., Univ. of California, Santa Cruz, CA 95064}

% \shortauthors{}

\begin{abstract}
We present the first scattered-light image of the debris disk around \target\ in $H$ band using the Gemini Planet Imager. \target\ is a $\sim 15$ Myr old A2IV star at a distance of $\sim 120$ pc in the Sco-Cen OB association. We detect the disk only in polarized light and place an upper limit on the peak total intensity. No point sources resembling exoplanets were identified. Compared to its mid-infrared thermal emission, the disk in scattered light shows similar orientation but different morphology. The scattered-light disk extends from $\sim 75$ to $\sim 210$ AU in the disk plane with roughly flat surface density. Our Monte Carlo radiative transfer model can well describe the observations with a model disk composed of a mixture of silicates and amorphous carbon. In addition to the obvious brightness asymmetry due to stronger forward scattering, we discover a weak brightness asymmetry along the major axis with the northeast side being 1.3 times brighter than the southwest side at a 3-$\sigma$ level.

\end{abstract}

\keywords{circumstellar matter --- infrared: stars --- stars: individual (HD 131835)}

\slugcomment{Draft \today}
\maketitle

%\clearpage % TEMP
%%%%%%%%%%%%%%%%%%%%%%%%% INTRODUCTION %%%%%%%%%%%%%%%%%%%%%%%%%%%%%%%%%%%%%%%
\section{INTRODUCTION} \label{sec:intro}
Debris disks are the remnant products of planet formation processes; within them, planetesimals collisionally evolve to generate dust disks visible in thermal emission and scattered light. The Gemini Planet Imager (GPI, \citealt{macintosh_etal14}) is one of the first high-contrast instruments equipped with the extreme adaptive optics (AO) specially designed for direct imaging and spectroscopy of exoplanets and debris disks. By resolving a debris disk, we can characterize the spatial distribution of dust grains, infer its dynamical history, and deduce the presence of unseen planets (see \citealt{wyatt_08} for a review). Unlike most debris disks which are gas depleted, carbon monoxide is detected in \target\ system \citep{moor_etal15}. Being a rare resolved debris disk with detected gas, \target\ serves as a unique target for studying the relationship between gas-dust physics and planetary science.

\target\ is an A2IV star \citep{houk82} in the Upper Centaurus Lupus (UCL) moving group \citep{rizzuto_etal11}, a subgroup of the Sco-Cen association. \target\ is $\sim$15 Myr old \citep{mamajek_etal02, pecaut_etal12} at a distance of 123 $^{+16}_{-13}$ pc \citep{vanleeuwen07}. Its IR emission was discovered by {\it Infrared Astronomical Satellite} \citep{moor_etal06}. \citet{chen_etal12} presented MIPS observations and showed that it is one of only four UCL/Lower Centaurus Crux A-type stars with $\rm L_{\rm IR} / L_\ast > 10^{-3}$, comparable to $\beta$ Pic. 

Recently, \citet{hung_etal15} resolved the debris disk around \target\ at 11.7 and 18.3 \um\ using Gemini/T-ReCS. A three-component dust disk model, composed of an unusually warm continuous disk and two rings, was able to simultaneously explain the spectral energy distribution (SED) and the mid-IR thermal images. Compared to recent disk studies with GPI, such as HD 115600 \citep{currie_etal15} and HD 106906 \citep{kalas_etal15}, \target\ is less inclined and the disk flux is more radially extended ($\sim 35$ to $\sim 400$ AU), allowing us to better study its morphological features. Here we report the first scattered light detection of dust surrounding \target\ in polarized light with GPI.

%%%%%%%%%%%%%%%%%%%%%%%%% Observation %%%%%%%%%%%%%%%%%%%%%%%%%%%%%%%%%%%%%%%%
\section{OBSERVATIONS \& DATA REDUCTION} \label{sec:obs_reduction}

\subsection{Observations} \label{sec:obs_reduction:observations}

We observed \target\ as one of our GPI Exoplanet Survey (GS-2015A-Q-500) campaign targets with GPI at the Gemini South Observatory, Cerro Pachon, Chile. On 2015 May 01, we obtained thirty-two 60-s exposures in $H$-band polarimetry mode \citep{perrin_etal15, hinkley_etal09} with waveplate angles of $0^\circ$, $22\fdg5$, $45^\circ$, and $67\fdg5$. On 2015 May 04, we obtained forty-one 60-s exposures in $H$-band spectroscopic mode. The observations in both modes were taken with the coronagraph and with the total field rotation $> 80^\circ$ at airmass $\leq 1.014$.

The GPI's field of view (FOV) is 2.7\arcsec\ square, with a scale of $14.166 \pm 0.007$ mas/pixel \citep[updated from][]{konopacky_etal14}. The radius of the $H$-band focal plane mask (FPM) is 0.123\arcsec. Because the star is behind the occulter in the coronagraphic mode, astrometric and photometric calibrations use satellite spots, which are diffracted starlight formed by a square pupil-plane grating \citep[e.g.,][]{wang_etal14, sivaramakrishnan_etal10, sivaramakrishnan_and_oppenheimer06}.

%%%%%%%%%%%%%%%%%%%%%%%%% Data Reduction %%%%%%%%%%%%%%%%%%%%%%%%%%%%%%%%%%%%%
\subsection{Data Reduction} \label{sec:obs_reduction:reduction}

The data were reduced using the GPI Data Reduction Pipeline (DRP, \citealt{perrin_etal14}). Polarimetry data were dark subtracted, corrected for flexure \citep{draper_etal14}, cleaned for correlated noise \citep{ingraham_etal14}, interpolated over bad pixels in the 2-D detector image, and assembled into data ``cubes'' with the third dimension comprising the two orthogonal polarization states. To get the Stokes cube, the images were divided by a low spatial frequency, polarized flat field \citep{millar-blanchaer_etal15}, interpolated over bad pixels in the 3-D datacubes, mitigated for systematics between the two orthogonal polarization channels via double differencing, subtracted for the instrumental polarization within each data cube \citep{wiktorowicz_etal14} using the average polarization fraction measured within the FPM, rotated to align the image orientations, and combined using singular value decomposition matrix inversion. 

The spectroscopic data were dark subtracted, flexure corrected and wavelength calibrated \citep{wolff_etal14} with an $H$-band Ar arc lamp taken right before the science sequence, interpolated over bad pixels in 2-D, assembled into a spectral datacube, interpolated over bad pixels in 3-D, and corrected for distortion.

%%%%%%%%%%%%%%%%%%%%%%%%% Calibration %%%%%%%%%%%%%%%%%%%%%%%%%%%%%%%%%%%%%
\subsection{Photometric Calibration} \label{sec:obs_reduction:calibration}

We perform photometric calibration on the polarimetry data by considering the satellite spot to star flux ratio $R$, the stellar flux $F_\star$ in physical units, and the average satellite spot flux $S$ in analog-to-digital unit (ADU) per coadd. The calibrated image data $D_f$ can be found by:  
\begin{equation}
D_f = D_i  \frac{R\,F_\star}{S},
\end{equation}
where $D_i$ is the image data in ADU coadd$^{-1}$. In GPI $H$-band, $R \sim 2 \times 10^{-4}$ \citep{wang_etal14}. We adopt $F_\star = 965 \pm 35$ mJy as the $H$-band flux of \target\ from 2MASS \citep{cutri_etal03}. We use an elongated aperture, similar to the shape of a running track, to perform aperture photometry on the satellite spots in polarimetry mode. We obtain a conversion factor of $1\ {\rm ADU\ coadd^{-1}\ s^{-1}} = (7.8\pm1.3) \times 10^{-4} {\rm\ mJy}$ with the uncertainty mostly stemming from measurement of $S$.

The photometric calibration in the spectral mode \citep{maire_etal14} is done using the Calibrated Datacube Extraction recipe\footnote{\url{http://docs.planetimager.org/pipeline/usage/tutorial_spectrophotometry.html}} via the GPI DRP based on the same principle. However, instead of using the broad band $F_\star$ and $S$ fluxes in the above equation, we replace them with the host star spectrum and the average satellite spot spectrum. For the stellar spectrum, we use the IDL Astrolib routine ccm\_unred.pro (based on \citealt{cardelli_etal89}) to apply $A(V)$ = 0.187 mag. reddening \citep{chen_etal12} to the \citet{kurucz_93} stellar atmosphere model with $T_{\rm eff}$ = 8770 K, $\log g$ = 4.0 and solar metallicity.

%%%%%%%%%%%%%%%%%%%%%%%%% Disk Morphology %%%%%%%%%%%%%%%%%%%%%%%%%%%%%%%%%%%%
\section{MORPHOLOGY OF THE SCATTERED-LIGHT DISK} \label{sec:morphology}
\begin{figure}
\epsscale{1.1}
\plotone{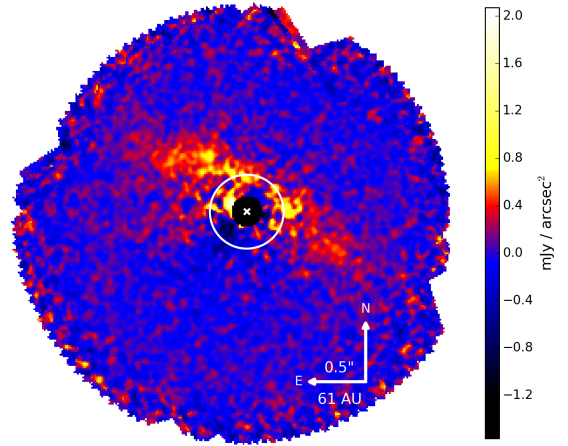}
\caption{GPI detection of dust-scattered star light around \target\ in \textit{H}-band tangentially polarized intensity. The image is smoothed by a 3-pixel FWHM Gaussian ($\sim$ the PSF size at 1.6 \um). The location of the star (white $\times$) and the FPM  (black circle) are marked. Residual instrumental polarization likely  affects the region within the white circle. The stronger forward scattering makes the front (NW) side of the disk more apparent. A weaker brightness asymmetry is detected along the major axis with the NE side being 1.3 times brighter than the SW side.}\label{fig:radial_pol}
\end{figure}

We resolve the debris disk around \target\ through polarimetric differential imaging (PDI). Figure~\ref{fig:radial_pol} shows the calibrated GPI $H$-band polarized intensity of \target\ in radial Stokes $Q_{\rm r}$ \citep{schmid_etal06}. In the sign convention adopted here, positive $Q_{\rm r}$ shows the tangentially polarized intensity while negative $Q_{\rm r}$ represents radially polarized intensity. The disk appears to be inclined with scattered light extending from $\sim$75--120~AU. By fitting the location of the flux peak along the major axis on each wing, we find no significant offset that is larger than 300 mas. If we assume an axisymmetric and radially smooth density structure, the projected eccentricity, $e$, along the major axis is consistent with zero, and $e > 0.2$ is rejected at 1-$\sigma$. Due to the non-detection on the southeast (SE) quadrant, the eccentricity along the minor axis is left unconstrained. The data are limited by instrumental polarization within the central $\sim 0.3\arcsec$ and by photon noise at larger radii.

Figure \ref{fig:radial_pol} shows brightness asymmetries along both the minor and major axes. The northwest (NW) side of the disk is significantly brighter than the SE side, which is undetected. This brightness asymmetry is likely due to light scattered in a preferential direction as seen in the case of HR 4796A \citep{perrin_etal15}. In addition, a weaker brightness asymmetry is present along the major axis. To quantify this brightness asymmetry, we use the best-fit geometric parameters found in \S \ref{sec:model} (except for setting the outer radius to be 180~AU due to the limited FOV) and consider the region exterior to $0.3\arcsec$ on the NW side of the major axis. Since single scattering by circumstellar dust is expected to produce linearly polarized light only in the $Q_{\rm r}$ polarization states, we measure the noise using Stokes $U_{\rm r}$ which corresponds to the linear polarization $45^\circ$ from $Q_{\rm r}$. The error at each angular separation in the $Q_{\rm r}$ image was estimated by measuring the standard deviation of the three-pixel wide annulus at the same separation in the Stokes $U_{\rm r}$ image. We find the northeast (NE) side of the disk is $1.30 \pm 0.09$ times brighter than the southwest (SW) side. In contrast, the thermal imaging shows the sides are equally bright, with a 30\% brightness asymmetry excluded at $> 3\sigma$ \citep{hung_etal15}.

%%%%%%%%%%%%%%%%%%%%%%%%% Disk Total Intensity and Exoplanet Search %%%%%%%%%%
\section{LIMITS ON DISK TOTAL INTENSITY AND POINT SOURCES} \label{sec:limits}

\subsection{Disk Total Intensity}
\label{totalintensity}
To subtract the stellar PSF from the total intensity polarimetry and spectroscopic images, we used a Python implementation of the Karhunen-Lo\`eve Image Projection (KLIP) algorithm \citep{soummer_etal12, pueyo_etal15}, pyKLIP \citep{wang_etal15} to perform PSF subtraction using angular differential imaging (ADI, \citealt{marois_etal06}). We divided the images into three annuli and four azimuthal subsections and ran KLIP over each zone, using an angular exclusion criterion of 5 degrees to select reference images. We used the first five KL basis vectors to estimate the PSF for each subsection. These parameters were selected by optimizing the throughput of an injected model disk. 

The disk is undetected in both Stokes $I$ and spectral data. The non-detection could be a result of the faintness of the disk as well as severe ADI self-subtraction. Even with the total field rotation being $> 80^\circ$, the radially extended geometry of the moderately inclined disk makes it particularly susceptible to the latter effect. To get an upper limit on the total intensity, we inject increasingly brighter model disks (discussed in \S~\ref{sec:model}) into the raw data and find when we can recover the disk after the PSF subtraction. The 3-$\sigma$ upper limit on the peak total intensity in the polarimetry data is 140 mJy arcsec$^{-2}$, giving a lower limit of the peak polarization fraction of 1\%. The spectral data gives a less constraining upper limit. These upper limits are larger than the total intensity of the best-fit model discussed in Section~\ref{sec:model}, thus demonstrating consistency between our empirical upper limits and our modeling. 

\subsection{Exoplanet Search} 
\label{planet}

We process our spectroscopic-mode data to optimize planet detection. We first subtract the PSF for each wavelength channel using the TLOCI code \citep{marois_etal14}, which assumes an input spectrum to optimize the subtraction while maximizing the signal-to-noise ratio (S/N) of an exoplanet of that specific spectral type. For our analysis, T8 and L8 spectra are chosen as the priors based on our experience to cover a wide range of DUSTY \citep{chabrier_etal00} and COND \citep{baraffe_etal03} exoplanets. The final data cube is then collapsed by a weighted mean, considering the input spectrum and the noise.

\begin{figure}
\epsscale{1.02}
\plotone{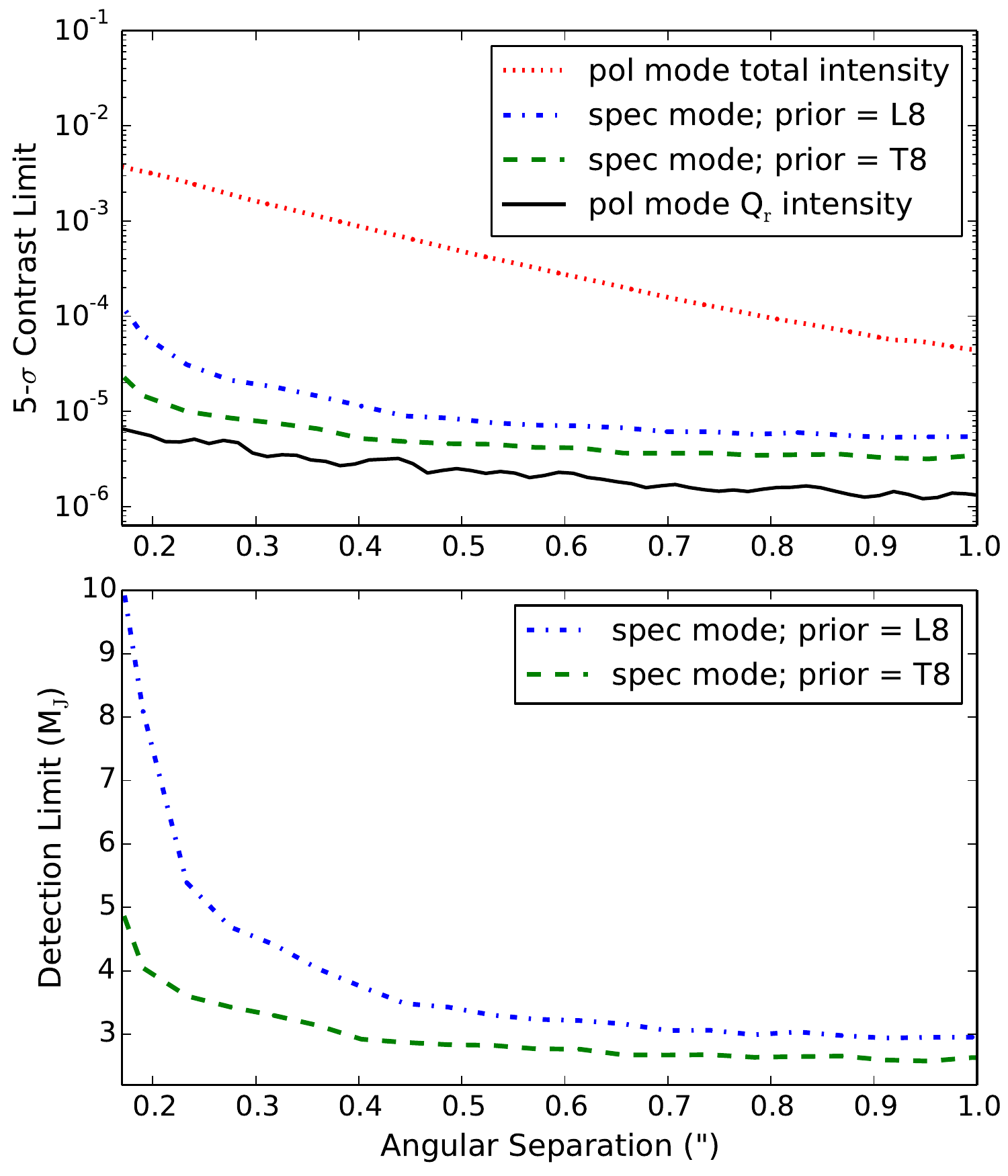}
\caption{Limits on point-sources detection from \textit{H}-band polarimetry and spectroscopic observations. Top: 5-$\sigma$ point-source contrast curves. Bottom: detection limit in terms of the mass of exoplanets. In the spectroscopic mode, prior spectra were used in the PSF subtraction so both contrast and mass limits are spectrum dependent.}\label{fig:spec_contrast}
\end{figure}

We search for point sources with planet-like spectra but detect none. To estimate our upper limits, we derive the TLOCI contrast curves by measuring the standard deviation of the pixel noise in each annulus of $\lambda/D$ width. These contrast curves are then transformed into exoplanet mass upper limits using the BT-settl models \citep{allard_etal12}. In our polarimetry mode observation, we derive the point source contrast curves by dividing the scatter (due to photon and read noise) at each annulus by the stellar flux, similar to how the contrast curves are derived in the spectroscopic mode. The contrast curves for the total and polarized intensities and mass limits derived from the spectral-mode observations are shown in Figure~\ref{fig:spec_contrast}. We reject objects with M $\gtrsim 3.5\ \rm M_J$ outside of $0.5 \arcsec$.

%%%%%%%%%%%%%%%%%%%%%%%%% Analysis %%%%%%%%%%%%%%%%%%%%%%%%%%%%%%%%%%%%%%%%%%%
\section{MODELING THE SCATTERED-LIGHT DISK} \label{sec:model}

We take a two-step approach to find a model that fits the SED and the GPI image. First, we use a geometric model to retrieve the structure of the scattered-light disk. Then, fixing the disk geometry, we search for a physical model that is built on the model proposed by \citet{hung_etal15} to get an estimate of the main dust properties associated with the polarized scattered light detection. In both steps, we exclude the central region within $0.3\arcsec$ due to uncorrected systematic errors from the instrumental polarization and cut out the region beyond 260~AU in the disk plane to reduce the number of pixels without detected disk signal. 

\begin{figure}
\epsscale{}
\plotone{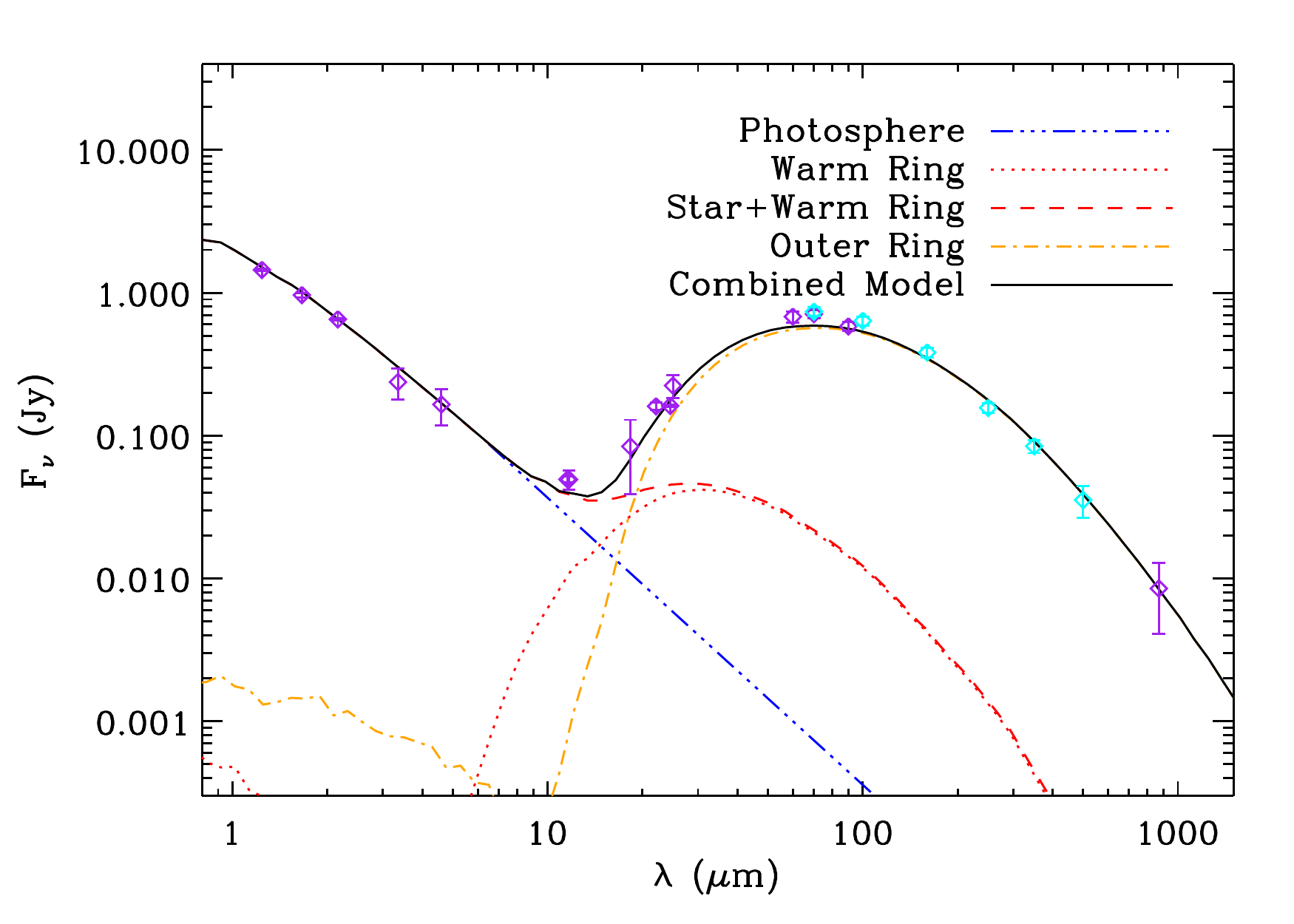}
\plotone{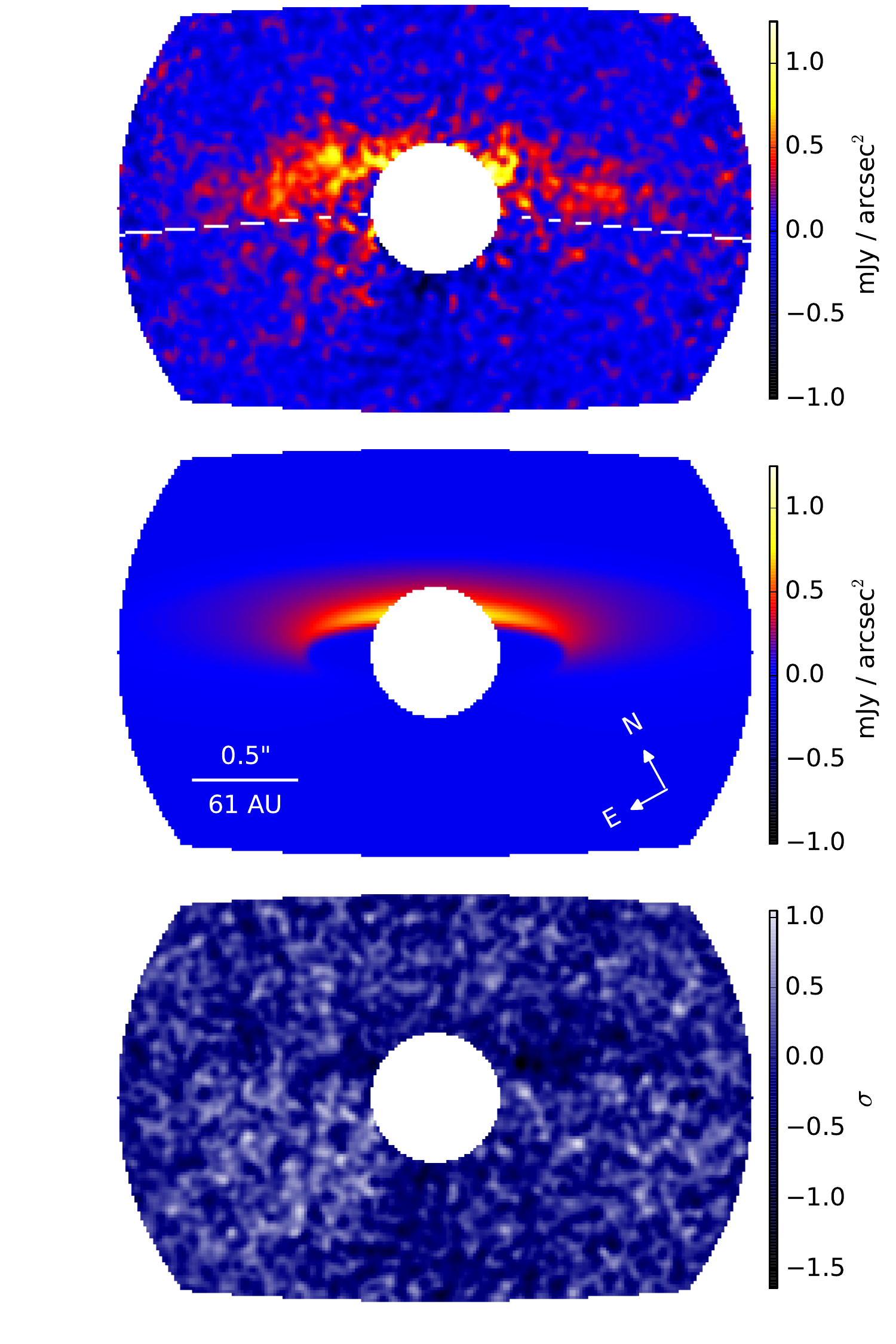}
\caption{Comparison of the best-fit MCFOST model to the 
observations. Top: SED with the best-fit model. The purple photometry values and references are summarized in \citet{hung_etal15}. The additional $Herschel$ points (cyan) are from \citet{moor_etal15}. Bottom: GPI polarized intensity data (Fig. \ref{fig:radial_pol}), best-fit model, and residual images viewed through the fitting mask. All displayed images were smoothed with a Gaussian with FWHM of 3 pixels. The geometric parameters of the MCFOST model were fixed to the values found by fitting the simple geometric model that only covers the azimuthal angles above the white dashed lines.}\label{fig:model}
\end{figure}

To measure the basic geometric properties of disk, we adopt a simple two-dimensional continuous disk model. Since the disk is not detected at all azimuths, we cut off $140^\circ$ symmetrically about the SE semi-minor axis in our model (white dashed lines in Figure ~\ref{fig:model}) to exclude the region with S/N $ \leq  1$. The model extends from the inner radius $r_{\rm in}$ to the outer radius $r_{\rm out}$ with the surface brightness varying only as $r^\alpha$. Along with the position angle $PA$ of the major axis and the inclination $i$, we use these five parameters to describe the geometric properties of the disk. We fit the data using the \texttt{emcee} python package \citep{foreman_etal13} based on the ensemble MCMC method of \citet{goodman_etal10}. After a burn-in period, we let the 100 walkers run for 1200 steps. The best-fit parameters and uncertainties listed in Table ~\ref{tab:model_parameters} are found by taking the median of the marginalized probability density distributions and finding the 1-$\sigma$ confidence intervals. The shapes of the posterior distributions are all single-peaked and approximately normal. Assuming that the disk is optically thin and the polarization fraction and the phase function do not depend on the stellocentric radius, the value of $\alpha$ implies a nearly flat surface density profile of $\Sigma(r) \propto r^{-0.3}$.

\begin{deluxetable}{ccc}
\tablewidth{0pc}
\tablecaption{Best-fit model parameters.\label{tab:model_parameters}}
\tablehead{\colhead{Parameter} &
\colhead{Value} &
\colhead{Units}}
\startdata
\cutinhead{Geometric parameters}
$PA$ & $61.4\pm0.4$ &  $\degr$ \\
$i$ & $75.1^{+0.8}_{-0.9}$ & $\degr$ \\
$r_{\rm in}$ & $75^{+2}_{-4}$ & AU\\
$r_{\rm out}$ & $210\pm10$ & AU\\
$\alpha$ & $-2.3^{+0.2}_{-0.1}$ &\\
\cutinhead{Hotter MCFOST model component}
$PA^*$ & 61.4 &  $\degr$ \\
$i^*$ & 75.1 & $\degr$ \\
$r^*$& 35 -- 400 & AU\\
$r_{\rm slope}^*$ & 0.5 & \\
$a^*$ & 0.03 -- 5 & \um\\
$a_{\rm slope}^*$ & $-3.5$ & \\
$M_{\rm dust}^*$ & $6.66\times 10^{-4}$ & $M_{\earth}$\\
${\rm composition}^*$ & amorphous carbon & \\
\cutinhead{Cooler MCFOST model component}
$PA^*$ & 61.4 &  $\degr$ \\
$i^*$ & 75.1 & $\degr$ \\
$r^*$& 75 -- 210 & AU\\
$r_{\rm slope}^*$ & $-0.3$ & \\
$a_{\rm min}$ & 1.56 & \um\\
$a_{\rm max}^*$ & 1. & mm\\
$a_{\rm slope}$ & $-3.46$ & \\
$M_{\rm dust}$ & $2.66\times 10^{-1}$ & $M_{\earth}$ \\
${\rm composition}^*$ & 50\% carbon + 50\% silicate & 
\enddata
\tablecomments{$^*$kept fixed}
\end{deluxetable}

Next, we use the radiative transfer code MCFOST \citep{pinte_etal06, pinte_etal09} to generate the polarized scattered light image and the SED. Using Mie theory, MCFOST self-consistently computes the absorption and scattering cross-sections as well as the scattering angle-dependent Mueller matrix, producing model images for all Stokes parameters. We assume a geometrically flat disk and start by modifying the two-component model from \citet{hung_etal15} to match the GPI image and the SED. We set the hotter extended component to have amorphous carbon \citep{li_etal97} as its composition based on the suggested high grain temperatures \citep{hung_etal15}. We keep all its parameters ($PA$, $i$, disk extent $r$, power-law index of the surface density $r_{\rm slope}$, grain size $a$, grain size distribution power-law index $a_{\rm slope}$, and mass of the dust $M_{\rm dust}$) fixed as listed in Table ~\ref{tab:model_parameters}. The cooler ring component has its geometry fixed to be the values found in the previous paragraph and composition set to be 50\% amorphous carbon and 50\% astro-silicate \citep{draine_etal84}. We have considered arguably simpler compositions (pure silicates and pure amorphous carbon) but they produce worse model fits. Adding water ice to the composition or porosity to the grains also leads to poorer model fits.

We simultaneously fit our MCFOST model to the SED and the scattered light disk at all azimuthal angles. In the fit, we weight the residuals from each pixel and broadband photometry point equally. In addition to fitting photometry points long-ward of 10 \um\ (summarized in \citealt{hung_etal15}), we include six new $Herschel$ points from \citet{moor_etal15}. With the disk geometry and dust composition fixed, the only free parameters are the size distribution and $M_{\rm dust}$. The SED can only place a lower limit on the maximum grain size so we adopt $a_{\rm max} = 1\ {\rm mm}$. This only leaves the minimum grain size $a_{\rm min}$, $a_{\rm slope}$, and $M_{\rm disk}$ of the cooler component to vary. To explore the parameter space, we use the genetic algorithm \citep{mathews_etal13} which ensures a fast convergence. 

The best-fit MCFOST model well reproduces the SED and the observed scattered-light disk with the reduced $\chi^2$ of 0.97. The fit is shown in Figure ~\ref{fig:model} and the parameters are listed in Table ~\ref{tab:model_parameters}. The best-fit model image fits the disk ansae well. The slight over-subtraction near the inner edge is not significant given the noise in these regions. Although the hotter component does not contribute significantly to the scattered-light image, this component is an important source for thermal emission. Our best-fit MCFOST model also roughly reproduces the extended thermal emission. We can further improve our model with detailed analysis, such as more complex dust composition and grain size distribution, but those are beyond the scope of this letter. Nonetheless, we set strong constraints on the disk properties, reproducing the surface brightness of the scattered-light disk with a model that was initially devised exclusively on thermal emission. We find a minimum grain size that is in reasonable agreement with the expected blowout size of 0.91 \um\ and the grain size power law index only slightly steeper than the canonical $a_{\rm slope} = -3.5$. Overall, the grains detected with GPI seems to follow the intuitive expectations of common disk properties.

%%%%%%%%%%%%%%%%%%%%%%%%% DISCUSSION %%%%%%%%%%%%%%%%%%%%%%%%%%%%%%%%%%%%%%%%%%%%
\section{DISCUSSION \& SUMMARY} \label{sec:discussion}

The mid-IR \citep{hung_etal15} and scattered-light images show different morphology. Unlike the continuous and extended thermal emission, the disk in scattered light has a cleared region inwards of $\sim 75$ AU. In other words, the scattered light disk starts at a radius that is twice as far away from the star compared to the disk in thermal emission. In polarized scattered light, we detect brightness asymmetries strongly along the minor axis and weakly along the major axis. The brightness asymmetry along the minor axis is likely due to asymmetry in the scattering phase function and is present in our best-fit model. The brightness asymmetry along the major axis could be the result of a dust density enhancement, azimuthal variation of grain compositions, or a projection effect of an eccentric disk if this 3-$\sigma$ feature is real. Since the mid-IR data do not show the asymmetry along the major axis, it suggests that the large grain population is more symmetrically arranged than the small grain population. The simulation done by \citet{wyatt_06} shows that dust that originates in the breakup of planetesimals trapped in resonance with a planet can have moderate-sized grains (a few \um\ to a few mm) distributed axisymmetrically but small grains (less than a few \um) exhibit trailing spiral structure that emanates from the resonant clumps. Therefore, the mismatched distributions of large and small grains can identify different forces acting on them and highlight potentially interesting dynamical interactions in the system.

\target\ is distinctive compared with the other Sco-Cen debris disks that have recently been imaged in scattered light: HIP 79977, HD 115600, HD 106906, and HD 110058 \citep{thalmann_etal13, currie_etal15, kalas_etal15, kasper_etal15}. Among those, \target\ has the largest inner radius in the scattered light-detected component and the most radially extended and nearly flat surface density profile (from 75 to 210 AU with $\Delta r / r \sim 1$). The other disks either have relatively narrow belts (HD 115600, HD 106906, and HD 110058) or have relatively sharp declines in brightness (HIP 79977). The extended and approximately flat surface density profile suggests the parent body belt of \target\ is likely to be extremely broad, much more so than any other debris disks imaged to date. This novel feature we found indicates the silicate component is not distributed in the form of one or two narrow rings as previously suggested by \citet{hung_etal15}. In addition, among all the Sco-Cen disks, \target\ is the only resolved disk with detected CO gas \citep{moor_etal15}, making it a unique and valuable target for studying gas-dust interactions. Besides the potential dynamical influence of undetected exoplanets, interactions between dust and gas could also play a significant role in clearing the dust in the inner disk and creating an eccentric ring \citep{lyra_etal12}.

Follow-up observations and detailed modeling are required to characterize the disk in detail. Deeper polarimetry observations are needed to confirm the NE-SW asymmetry. Detection of the disk in total intensity can set a firm constraint on the grain shape and porosity by providing the information on the fractional polarization as a function of scattering angle. Multicolor observations can further constrain the grain composition. Since \target\ is located in the southern sky, GPI, SPHERE (Spectro-Polarimetric High-contrast Exoplanet Research), and ALMA (Atacama Large Millimeter/submillimeter Array) are powerful instruments/facility for conducting follow-up observations.

%%%%%%%%%%%%%%%%%%%%%%%%% ACKNOWLEDGEMENTS %%%%%%%%%%%%%%%%%%%%%%%%%%%%%%%%%%%
\acknowledgements

This research was supported in part by NASA cooperative agreements NNX15AD95G, NNX11AD21G, and NNX14AJ80G, NSF AST-113718, AST-0909188, AST-1411868, AST-1413718, and DE-AC52-07NA27344, and the U.S. Department of Energy by Lawrence Livermore National Laboratory under Contract DE-AC52-07NA27344. Work by L.-W. Hung and A. Greenbaum is supported by the National Science Foundation Graduate Research Fellowship DGE-1144087 and DGE-1232825. We acknowledge the Service Commun de Calcul Intensif de l'Observatoire de Grenoble (SCCI) for computations on the super-computer funded by ANR (contracts ANR-07-BLAN-0221, ANR-2010- JCJC-0504-01 and ANR-2010-JCJC-0501-01) and the European Commission's 7th Framework Program (contract PERG06- GA-2009-256513). Based on observations obtained at the Gemini Observatory, which is operated by the Association of Universities for Research in Astronomy, Inc., under a cooperative agreement with the NSF on behalf of the Gemini partnership: the National Science Foundation (United States), the National Research Council (Canada), CONICYT (Chile), the Australian Research Council (Australia), Minist\'{e}rio da Ci\^{e}ncia, Tecnologia e Inova\c{c}\~{a}o (Brazil) and Ministerio de Ciencia, Tecnolog\'{i}a e Innovaci\'{o}n Productiva (Argentina).

%%%%%%%%%%%%%%%%%%%%%%%%% REFERENCES %%%%%%%%%%%%%%%%%%%%%%%%%%%%%%%%%%%%%%%%%
% references
%\bibliography{literature}

\newcommand{\noopsort}[1]{}

\end{document}